\documentclass[aps,prl,twocolumn,superscriptaddress,groupedaddress]{revtex4}  
\usepackage{graphicx}
\usepackage{dcolumn}
\usepackage{amsmath,amssymb,epsfig,bm,amsfonts,yfonts,bbm,mathrsfs}

\begin{document}

\title{Thermal excitation spectrum from entanglement in an expanding quantum string}
\author{J\"{u}rgen Berges}
\email{berges@thphys.uni-heidelberg.de}
\affiliation{Institut f\"{u}r Theoretische Physik, Universit\"{a}t Heidelberg, Philosophenweg 16, 69120 Heidelberg, Germany}
\author{Stefan Floerchinger}
\email{floerchinger@thphys.uni-heidelberg.de}
\affiliation{Institut f\"{u}r Theoretische Physik, Universit\"{a}t Heidelberg, Philosophenweg 16, 69120 Heidelberg, Germany}
\author{Raju Venugopalan}
\email{raju@bnl.gov}
\affiliation{Physics Department, Brookhaven National Laboratory, Bldg. 510A, Upton, NY 11973, USA}


\begin{abstract}
A surprising result in $e^+ e^-$ collisions is that the particle spectra from the string formed between the expanding quark-antiquark pair have thermal properties even though scatterings appear not to be frequent enough to explain this. We address this problem by considering the finite observable interval of a relativistic quantum string in terms of its reduced density operator by tracing over the complement region. We show how quantum entanglement in the presence of a horizon in spacetime for the causal transfer of information leads locally to a reduced mixed-state density operator. For very early proper time $\tau$, we show that the entanglement entropy becomes extensive and scales with the rapidity. At these early times, the reduced density operator is of thermal form, with an entanglement temperature $T_\tau=\hbar/(2\pi k_B \tau)$, even in the absence of any scatterings. 
\end{abstract}

\maketitle

\paragraph{Introduction.}

A longstanding puzzle in $e^+ e^-$ collisions is that the hadron spectra measured appear thermal with features that can be characterized in terms of a common temperature~\cite{Becattini:1995if,Becattini:2009fv,Becattini:2010sk, Castorina:2007eb, Andronic:2008ev, Fischer:2016zzs}. The apparent thermal origin of the multiparticle production is surprising because scatterings appear not to be frequent enough for thermalization to occur and therefore demands an alternative explanation~\cite{Elze:1994hj, Elze:1994qa, Bialas:1999zg, Castorina:2007eb, Akkelin:2016rhm}. 

We argue in this letter that this apparent thermalization is an intrinsically quantum phenomenon arising from the entanglement between observable and unobservable regions in an expanding string. The observable region is described in terms of a reduced density operator by tracing over the complement region, which is bounded by the Minkowski spacetime horizon for the causal transfer of information. 
We show that the entanglement of the quantum vacuum accross this horizon leads to dramatic macroscopic quantum effects. In particular, for very early proper time $\tau$, we discover that the entanglement entropy is extensive and scales with the rapidity. At these early times, a conformal symmetry emerges for the expanding system and the entanglement generates a reduced density matrix of thermal form, with the temperature 
\begin{equation}
T_\tau=\frac{\hbar}{2\pi k_B \tau}  \, ,
\label{eq:quantum temperature}
\end{equation}
even in the absence of any scatterings.

Our results establish a novel class of horizon phenomena in quantum field theory, featuring an instantaneous thermal excitation spectrum from a vacuum pure state. In contrast to the well-known example of an event horizon in the vicinity of a black hole, which leads to Hawking radiation, or the related Unruh temperature for a class of accelerated observers, our setting does not involve acceleration and it is non-stationary~\cite{Unruh:1976db}. Specifically, the Unruh acceleration $a$ of an observer in the Rindler-wedge of Minkowski spacetime at a spatial position $x=c^2/a$ generates a space-dependent temperature $T_x=\hbar c/(2\pi k_B x)$, while the time-dependent temperature (\ref{eq:quantum temperature}) applies to the initial stages in the forward light cone with crucial applications to $e^+ e^-$ but also hadron-hadron collisions.

\paragraph{Model of expanding strings.}
Models that describe  $e^+ e^-$ collisions successfully~\cite{Andersson:1983ia, Catani:1991hj} rely on the Schwinger mechanism of particle production in 1+1-dimensional quantum electrodynamics (QED); a recent comprehensive discussion of the  difficulties presented by thermal-like spectra in such models can be found in ~\cite{Fischer:2016zzs}. We will work within this Schwinger model framework to treat the dynamics of the expanding string formed  between the relativistic quark-antiquark pair. We choose the coordinate system such that the trajectories are in natural units $z=\pm t$, $x=y=0$, and we assume that the strings are essentially confined to the $z$-direction. Bjorken coordinates are convenient, with $z=\tau \sinh(\eta)$ and $t = \tau \cosh(\eta)$ with rapidity $\eta$ and proper time $\tau=\sqrt{t^2-z^2}$. In these coordinates, the Minkowski space metric in the confined space can be expressed as $ds^2=-d\tau^2+\tau^2 d\eta^2$. 

The Schwinger model is particularly simple for a single massless Dirac fermion. In this case, it can be bosonized  to a free massive scalar theory with the action~\cite{Frishman:2010zz}
\begin{equation}
S = \int d^2 x \sqrt{g} \left\{ - \frac{1}{2}g^{\mu\nu}\partial_\mu \phi \partial_\nu \phi - \frac{1}{2}M^2 \phi^2 \right\}.
\label{eq:bosonizedLagrangian}
\end{equation}
For convenience, we have employed general coordinates with the two-dimensional metric $g_{\mu\nu}$. The Schwinger model bosons $\phi$ correspond to dipoles that are quadratic in the original fermion field.  Their mass is proportional to the U(1) charge, $M=q/\sqrt{\pi}$; likewise, the string tension satisfies $\sigma=q^2/2$. Bosonization also works for a nonvanishing fermion mass $m$ but we will not consider that case here. 

\paragraph{Dynamics of expansion.}
For the bosonized Schwinger model, a solution corresponding to an expanding string stretched between two external quarks on their lightcones is found as a rapidity invariant solution to the equation of motion, $\partial_\tau^2 \bar \phi + \partial_\tau \bar \phi/\tau + M^2 \bar\phi=0$. The boundary condition for $\tau\to 0_+$ is fixed by the requirement that the electric field $E=q \phi/\sqrt{\pi}$ approaches the U(1) charge of the external quarks $E\to q_\text{e}$. This gives  $\bar\phi(\tau)\to \sqrt{\pi}q_\text{e}/q$, and with this boundary condition one finds $\bar\phi(\tau)=\sqrt{\pi}(q_\text{e}/q) J_0(M\tau)$. The oscillations  in this solution are related to multiple string breaking~\cite{Hebenstreit:2013baa}. 

The solution $\bar \phi = \langle \phi(x) \rangle$ should be understood as a field expectation value, or equivalently, as a coherent field. Further information is contained in correlation functions of the fields $\phi(x)$ and their conjugate momentum fields $\pi(x)$, which specify a density matrix $\rho$ at some initial time or on an  appropriate Cauchy hypersurface. Because the action \eqref{eq:bosonizedLagrangian} is quadratic in the field $\phi$, we also assume that this density matrix is of Gaussian form. 
Gaussian density matrices are fixed entirely in terms of the expectation values and connected two-point correlation functions.

\paragraph{Entanglement entropy of a rapidity interval.}

\begin{figure}
	\centering
	\includegraphics[width=0.48\textwidth]{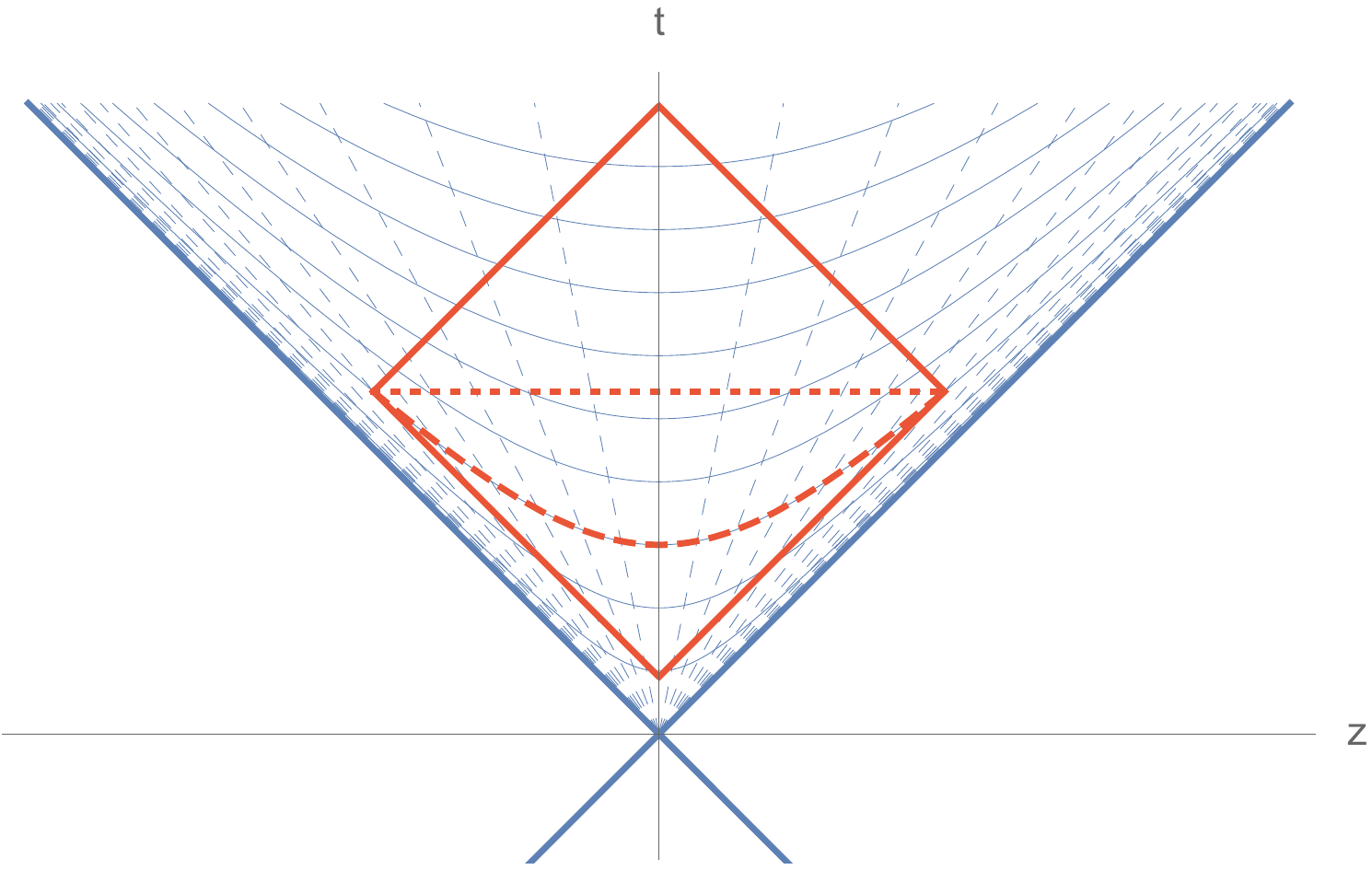}
	\caption{Bjorken coordinates and causal development of a rapidity interval $(-\Delta\eta/2,\Delta\eta/2)$ at fixed proper time $\tau$. The dashed red line corresponds to region $A$ while the complement region $B$ is composed of the rapidity intervals $(-\infty, -\Delta\eta/2)$ and $(\Delta\eta/2, \infty)$ at fixed Bjorken time $\tau$. For $\Delta\eta\to \infty$ the causal development region approaches the lightcone.}
	\label{fig1}
\end{figure}

To discuss processes, such as the formation of hadrons or resonances during the relativistic expansion of the string, we will assume that the dynamics is local in a space-like region $A$ within the future light cone of the spacetime instant of an $e^+ e^-$ collision. This local dynamics can be described by the reduced density matrix $\rho_A$, defined as the trace of $\rho$ over the complement region $B$:
\begin{equation}
\rho_A = \text{Tr}_B \rho\,.
\end{equation}
If the fields $\phi$ in the regions $A$ and $B$ are entangled, the reduced density matrix $\rho_A$ is of mixed form, even if the full density matrix $\rho$ is pure. The degree of entanglement, and therefore the deviation of $\rho_A$ from a pure state, can be characterized in terms of the entanglement entropy, 
\begin{equation}
S_A = - \text{Tr} \left\{ \rho_A \ln(\rho_A) \right\}\,.
\end{equation}

For most problems in quantum field theory, the determination of the reduced density matrix $\rho_A$, as well as of the entanglement entropy $S_A$, are formidable tasks. Results are currently known for free field theories in static situations \cite{Casini:2009sr}, or conformal field theories \cite{Holzhey:1994we, Calabrese:2004eu, Calabrese:2009qy}. For a discussion of entanglement entropy in the 't Hooft model, see \cite{Goykhman:2015sga}. The treatment of entanglement in nonequilibrium situations is especially difficult. In a companion paper \cite{BFV2}, we develop real-time techniques for (relative) entanglement entropies of general Gaussian states in quantum field theory. 

In the following, we will take advantage of the fact that if the density matrix $\rho$ is Gaussian (in the field theoretic sense) then this is also the case for the reduced density matrix $\rho_A$ of any equilibrium or nonequilibrium state. The entanglement entropy $S_A$ in this case is then given by \cite{BFV2}
\begin{equation}
S_A = \frac{1}{2} \text{Tr}_A\left\{ D \ln(D^2)\right\},
\label{eq:EntEntropyGaussian}
\end{equation}
where the operator trace is restricted to the region $A$ and the matrix $D$ consists of connected correlation functions. For the example of the bosonized Schwinger model with field $\phi$ and conjugate momentum field $\pi$, we obtain
\begin{equation}
D(x,y) = 
\begin{pmatrix}
-i \langle \phi(x) \pi(y) \rangle_c   &&  i \langle \phi(x) \phi(y) \rangle_c  \\
-i \langle \pi(x) \pi(y) \rangle_c &&  i \langle \pi(x) \phi(y) \rangle_c
\end{pmatrix}.
\label{eq:DMatrix}
\end{equation}
The expectation value $\langle \cdots \rangle$ in eq.\ \eqref{eq:DMatrix} can equivalently be taken with respect to the full density matrix $\rho$ or the reduced density matrix $\rho_A$.
For the specific case of the expanding relativistic string, we will compute the trace in eq.\ (\ref{eq:EntEntropyGaussian}) by taking $A$ to be the rapidity interval $(-\Delta\eta/2,\Delta\eta/2)$ at fixed Bjorken time $\tau$, corresponding to the dashed red line in Fig.~\ref{fig1}. The complement region $B$ corresponds to the sum of the rapidity intervals $(-\infty, - \Delta\eta/2)$ and $(\Delta\eta/2, \infty)$ at fixed Bjorken time $\tau$. Note that any process (e. g. a measurement) within the causal development region delimited by the solid red line in Fig.~\ref{fig1} is by reasons of causality at most sensitive to the region $A$ while the density matrix in the complement region $B$ cannot affect  such processes. 

 Because the field expectation values $\bar \phi(x)=\langle \phi(x)\rangle$ and $\bar \pi(x)=\langle \pi(x) \rangle$ do not enter \eqref{eq:EntEntropyGaussian} and \eqref{eq:DMatrix}, the entanglement entropy for an expanding string described by the massless Schwinger model (corresponding to a coherent state specified by $\bar \phi(x)$ and $\bar \pi(x)$) is the same as the one of the vacuum (which is a coherent state with vanishing field expectation values). Moreover, the entropy does not change under unitary evolution with the boundary points kept fixed. It can therefore equivalently be evaluated in the interval $(-\Delta z / 2, \Delta z/2)$ with $\Delta z = 2 \tau \,\text{sinh}(\Delta\eta/2)$ at fixed time $t=\tau \cosh(\Delta\eta/2)$, corresponding to the dotted red line in Fig.\ \ref{fig1}. 

Following this identification, our computation of the entanglement entropy reduces to an eigenvalue problem for which one can employ a discrete basis involving Fourier expansion on a finite interval. The trace in (\ref{eq:EntEntropyGaussian}) then involves contour integrals in momentum space, where the nonvanishing contributions arise from branch cuts~\cite{BFV2}. We emphasize that these contributions would be missing if naive discretizations with periodic boundary conditions -- as are often employed in calculations of nonequilibrium dynamics -- are assumed.    

\begin{figure}
	\centering
	\includegraphics[width=0.4\textwidth]{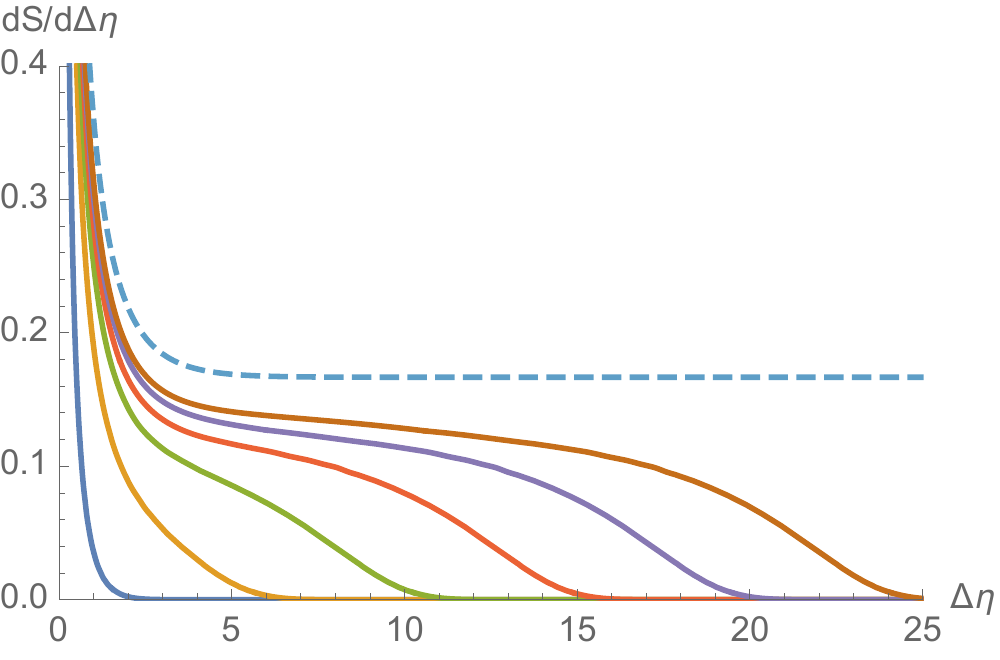}
	\caption{Entanglement entropy density $dS/d\Delta\eta$ as a function of the rapidity interval $\Delta\eta$ in the bosonized massless Schwinger model for the case of a massive scalar boson. The curves (from left to right)  correspond to $M\tau=1$, $M\tau=10^{-1}$, $M\tau=10^{-2}$, $M\tau=10^{-3}$, $M\tau=10^{-4}$, and $M\tau=10^{-5}$. For very early times, $M\tau\to 0$, a plateau forms corresponding to the asymptotic conformal case $dS/d\Delta\eta\to 1/6$ (dashed line).}
	\label{fig2}
\end{figure}

We first consider the massless limit $M=0$. Here, our result for the entanglement entropy corresponds to the  expression obtained previously for conformal field theories \cite{Holzhey:1994we, Calabrese:2004eu, Calabrese:2009qy}. In our case, the result can be expressed as 
\begin{equation}
S(\tau, \Delta\eta) = \frac{c}{3} \ln \left( 2\tau \,\text{sinh}(\Delta\eta/2)/\epsilon\right)+\text{constant},
\label{eq:EntEntConformalBjorken}
\end{equation}
where $c$ is the conformal central charge and $\epsilon$ is a length scale corresponding to an ultraviolet  cutoff. The additive constant is not universal, but the derivative of $S$ with respect to the interval length $\Delta z$ is. This implies
$\tau \partial S/\partial\tau  = c/3$ and $\partial S/\partial\Delta\eta = (c/6)\, \text{coth}(\Delta\eta/2)$. For a large rapidity interval $\Delta\eta\gg 1$, one has $S = (c/6) [ \Delta\eta +2 \ln (\tau)]+\text{const}$.
This demonstrates the existence of  a time independent piece of the entanglement entropy that is extensive in rapidity and a  $\Delta\eta$-independent piece that grows logarithmically with the proper time. 

Turning now to the non-conformal free massive scalars of the Schwinger model, the universal part of the entanglement entropy behaves just as in the conformal case for $M\Delta z \ll 1$ and decays for $M \Delta z \gg 1$ \cite{Casini:2009sr}. We are particularly interested in the dependence on $\Delta\eta$:
\begin{equation}
\begin{split}
& \frac{\partial}{\partial \Delta \eta}S(\tau, \Delta\eta) =  \frac{\partial S}{\partial \ln \Delta z} \frac{\partial \ln \Delta z}{\partial \Delta\eta} \\
& =  c_E\left(2M\tau \sinh(\Delta\eta/2)\right)  \coth(\Delta\eta/2)/2\,,
\end{split}
\end{equation}
where $c_E\left(M\Delta z\right) = \Delta z {\partial S}/{\partial \Delta z}$ is the entanglement entropy $c$-function for a massive scalar field. Taking the  conformal limit, one obtains, as anticipated, that $c_E(0)=c/3=1/3$. For large values of the argument, the function has the exponential decay form $c_E(x)\to x K_1(2x)/4$. 

Employing the numerically known expression for $c_E(x)$~\cite{Casini:2009sr}, we can compute $dS/d\Delta\eta$ for the bosonized massless Schwinger model. The result is shown in Fig.\ \ref{fig2} for different values of $M\tau$. We first note that for short times $\tau$ (compared to the mass $M$ or string tension $\sigma$), there is substantial entanglement over rapidity intervals $\Delta\eta={\cal O}(1)$. For intermediate values of $M\tau$ and $\Delta\eta$, one observes that $\partial S/\partial \Delta\eta$ approaches a plateau at $1/6$ as a function of $\Delta\eta$ at early times. We also observe that it decays  both for very large $\Delta\eta$ and for later times $\tau$. 

Remarkably, because the limit of very early times for expanding strings is equivalent to that of small mass $M$ or string tension $\sigma$, the conformal limit is recovered at time scales where the quasiparticle constituents in the Schwinger model are  effectively non-interacting.  This is because the dynamics at very early times is dominated by the expansion with ``Hubble rate'' $H=1/\tau \gg M=q/\sqrt{\pi}=\sqrt{2\sigma/\pi}$. (Note that in 1+1 dimensions the charge $q$ has dimensions of energy.) 

\begin{figure}
	\centering
	\includegraphics[width=0.4\textwidth]{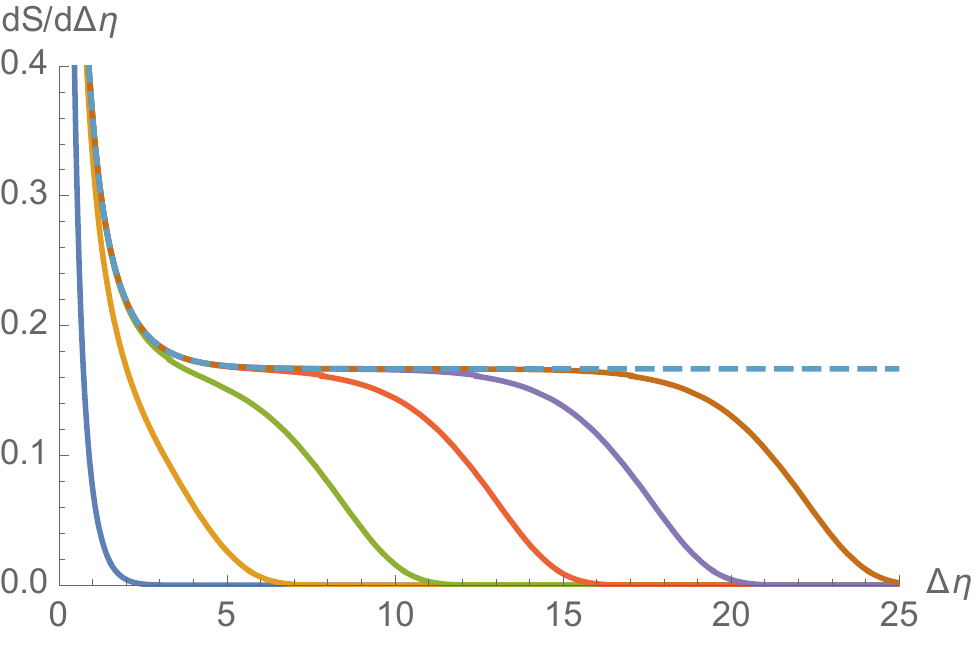}
	\caption{Entanglement entropy density $dS/d\Delta\eta$ as a function of the rapidity interval $\Delta\eta$ for a free massive Dirac fermion field. The curves (from left to right) correspond to $M\tau=1$, $M\tau=10^{-1}$, $M\tau=10^{-2}$, $M\tau=10^{-3}$, $M\tau=10^{-4}$, and $M\tau=10^{-5}$. In the limit of early time, $M\tau\to 0$, a plateau forms at the asymptotic value $dS/d\Delta\eta\to 1/6$, corresponding to the conformal case (dashed line).}
	\label{fig3}
\end{figure}

Since the entanglement entropy $c$-function in the conformal limit is identical for real massless scalar bosons and for massless Dirac fermions, this indicates that the conformal limit is consistently described with or without bosonization. One may also determine the entanglement entropy of free massive Dirac fermions using similar manipulations as described above and the corresponding $c$-function as given in \cite{Casini:2009sr}. The result is shown in Fig.\ \ref{fig3}. The approach to the universal plateau at $dS/d\Delta\eta=1/6$ for $M\tau\to 0$ is even faster for the free fermion case.

It is interesting to consider further the implications of these results. For a realistic description of hadronization, entanglement can well be the most relevant source of entropy. Moreover, if the processes are essentially dominated by the plateau in entanglement entropy density at early times, one would infer that the relation
\begin{equation}
\frac{\partial S}{\partial \eta} = \frac{\partial S}{\partial \Delta \eta} = \frac{c}{6} \,,
\end{equation}
in terms of the conformal charge $c$, is universal.

With this in mind, we can attempt to estimate the relevant conformal charge for the `t Hooft model~\cite{tHooft:1974pnl} which is the QCD analogue of two-dimensional QED.
There are no dynamical gluons in two dimensions and the residual effect of gauge interactions is negligible at early times when, as noted, the expansion rate is larger than the gauge coupling. We expect that $N_c\times N_f$ quarks with mass $m$ play a role (with $N_c=N_f=3$ for realistic energies); for $1/\tau \gg m$, they are effectively massless. Beyond the strict two dimensional approximation, the QCD string has two additional degrees of freedom corresponding to its fluctuating transverse coordinates. In the Nambu-Goto action, these correspond to two massless scalar fields~\cite{Brandt:2016xsp}. Taking them into account as free scalars would lead to $c=N_c\times N_f+2=11$. 

Unfortunately, extracting the entanglement entropy from experiment is not straightforward. What is measured in $e^+ e^-$ is the number of charged particles per unit rapidity $dN_\text{ch}/d\eta$, where rapidity here is defined with respect to the thrust axis. When evaluated around mid-rapidity this observable increases with the collision energy, albeit only logarithmically. Typical values for collision energies between $\sqrt{s}=14$ GeV and $\sqrt{s}=206$ GeV are in the range $dN_\text{ch}/d\eta \approx 2 - 4$, see \cite{GrosseOetringhaus:2009kz} for a review. The entropy per particle $S/N$ can be estimated for a hadron resonance gas in thermal equilibrium, with a typical value $S/N_\text{ch}=7.2$ \cite{Andronic:2012ut,Pal:2003rz} giving $dS/ d\eta \approx 14 - 28$ for the above mentioned energy range. This is significantly higher than our simple estimate. However, the entanglement entropy includes many-body correlations that will lower its value relative to the entropy estimated from single particle inclusive distributions. 

\paragraph{Thermal entanglement entropy.}
A remarkable consequence of our results is that the excitation spectrum for sufficiently early Bjorken time $\tau$ exhibits thermal properties even though the quasiparticles of the system are non-interacting at these early times! We start by observing that the entanglement entropy \eqref{eq:EntEntConformalBjorken} of the Schwinger model for $M \tau \rightarrow 0$ closely resembles the entanglement entropy of a conformal field theory at finite temperature in an interval of length $l$ \cite{Calabrese:2004eu}:
\begin{equation}
S(T,l) = \frac{c}{3} \ln\left( \frac{1}{\pi T \epsilon}  \sinh(\pi l T) \right)+\text{const}\,.
\end{equation} 
In fact, the expressions agree if one sets $l=\tau \Delta\eta$ (in the spirit of a ``direct translation'' from flat space to the expanding geometry with metric $ds^2=-d\tau^2+\tau^2 d\eta^2$) and identifies $T={1}/(2\pi \tau)$ or (\ref{eq:quantum temperature}) in conventional units. 
	
The nature of the thermal-like state with time-dependent temperature can be made more precise for conformal fields. Towards this end, we consider a conformal field theory in a spacetime region that is bounded by two light cones, represented for instance by the diamond shaped region enclosed by the solid red lines in Fig.\ \ref{fig1}. On any hypersurface $\Sigma$ with its boundary on the intersection of the two light cones, one can express the reduced density matrix as
\begin{equation}
\rho_A = \frac{1}{Z_A} e^{-K}, \quad\quad\quad Z_A = \text{Tr} \; e^{-K},
\label{eq:stat-factor}
\end{equation}
where the so-called modular or entanglement Hamiltonian is a local expression given by  \cite{Casini:2011kv, Arias:2016nip} (see also \cite{Candelas:1978gf})
\begin{equation}
K = \int_\Sigma d\Sigma^\mu \,  \xi^\nu(x) \, T_{\mu\nu}(x).
\label{eq:LocalModularHamiltonian}
\end{equation}
Here $T_{\mu\nu}(x)$ is the energy-momentum tensor and $\xi^\nu(x)$ is a vector field that can be written as 
\begin{equation}
\begin{split}
\xi^\mu(x) & =  \tfrac{2\pi}{(k-p)^2} [ (k-x)^\mu (x-p)(k-p) + (x-p)^\mu \\
& \times (k-x)(k-p) - (k-p)^\mu (x-p)(k-x)]\,,
\end{split}
\end{equation}
where $k$ represents the end point of the future light cone and $p$ the starting point of the past light cone. We note that \eqref{eq:LocalModularHamiltonian} is of the same form as a density matrix of a local thermal equilibrium state with $\beta^\mu = u^\mu/T = \xi^\mu$, the four-vector formed by the inverse of temperature $T$ and fluid velocity $u^\mu$ \cite{beta}. The vector $\xi^\mu$ vanishes on the boundary of the region enclosed by the two light cones, corresponding formally to an infinite temperature. The interpretation of $\xi^\mu$ as an inverse temperature vector can be made more precise in terms of the relative entropy for states that deviate from the vacuum state considered. More specifically, the probability to find a localized fluctuation with four-momentum $p^\mu$ in a field is given by a Boltzmann thermal weight involving $\xi^\mu p_\mu$ \cite{Arias:2016nip}.

As a consequence of causality, the production of hadrons and resonances as excitations of the vacuum state of the expanding string is confined to the region formed by the intersection of two light cones. In the 
$e^+ e^-$ case, the past light cone originates at the collision point $p=0$, while the future light cone is determined by the produced particles. Taking the corresponding end point $k$  to be timelike with $-k^2\to \infty$, but keeping $x^\mu$ finite, leads to $\xi^\mu=2\pi x^\mu$. In the Bjorken coordinates of the expanding string, this corresponds to a fluid velocity $u^\mu$ pointing in $\tau$ direction and the time-dependent temperature (\ref{eq:quantum temperature}). The argument above is not restricted to dynamics in 1+1 dimensions but holds in general for the double cone geometry and for conformal fields.

The modular Hamiltonian is known within the double cone region only for a conformal field theory. Moreover, $\xi^\mu$ is a conformal Killing vector; the local equilibrium interpretation can therefore only hold for conformal fields. However, as we argued, the real-time dynamics of the Schwinger model at early times is nearly conformal suggesting that our equilibrium picture of $\xi^\mu$ representing the inverse temperature is robust at these early times. If the system disintegrates shortly thereafter, particles will be produced according to 
the distribution (\ref{eq:stat-factor}), thereby providing an alternative explanation of statistical hadronization to be a consequence of quantum entanglement. If the system doesn't fall apart quickly enough, it is still conceivable that particle production is substantially affected by the quantum entanglement described here. However in this case, for non-conformal fields, the modular Hamiltonian will in general contain additional nonlocal terms with observable corrections.

\paragraph{Conclusions.} The entanglement of the observable and unobservable regions of an expanding quantum field theoretical string formed in $e^+ e^-$ collisions can lead to dramatic consequences. 
At very early times, the reduced density matrix corresponds to thermal excitations of a conformal field theory.  More generally, it is of mixed state form with a sizable extensive entanglement entropy per unit rapidity. These results suggest that the long-standing experimental puzzle of why statistical hadronization models in $e^+ e^-$ collisions are successful may be a consequence of quantum entanglement as opposed to multiparticle scatterings. Since stringlike dynamics is a ubiquitous feature of hadron-hadron collisions as well, it is also plausible that quantum entanglement may also play a role in the apparently early thermalization observed at RHIC and the LHC.

Our findings open up a wide range of possible new nonequilibrium applications where similar horizon phenomena from entanglement in quantum field theory can play an important role. While here we formulate our results for relativistic dynamics encountered in particle collider experiments, related questions apply also to non-relativistic systems with a unitary time evolution and a sound horizon. For instance, it would be very interesting to investigate the remarkable observation of thermal-like states immediately after a sudden quench in one-dimensional split Bose condensate experiments in view of our findings~\cite{coldgas}. Finally, the formalism we have developed further in \cite{BFV2} can be extended beyond 1+1-dimensions and applied to study the role of quantum entanglement in the spacetime evolution of, e.g., strong color fields in 3+1 dimensions~\cite{Kovner:2015hga,Kharzeev:2017qzs,Martens:2017cvj,Shuryak:2017phz}.

\paragraph{Acknowledgments.} This work is part of and supported by the DFG Collaborative Research Centre ``SFB 1225 (ISOQUANT)''. R.~V.'s research is supported by the U.\ S.\ Department of Energy Office of Science, Office of Nuclear Physics, under contracts No.\ DE-SC0012704. We would like to thank A.~Andronic, D.~Kharzeev and K.~Reygers for useful discussions. R.~V.\ would like to thank ITP Heidelberg and the Alexander von Humboldt Foundation for support, and ITP Heidelberg for their kind hospitality.


\vspace{-0.2cm}
\begin{thebibliography}{99}
%
\bibitem{Becattini:1995if} 
  F.~Becattini,
  Z.\ Phys.\ C {\bf 69}, no. 3, 485 (1996).

\bibitem{Castorina:2007eb} 
  P.~Castorina, D.~Kharzeev and H.~Satz,
  Eur.\ Phys.\ J.\ C {\bf 52}, 187 (2007).

\bibitem{Andronic:2008ev} 
  A.~Andronic, F.~Beutler, P.~Braun-Munzinger, K.~Redlich and J.~Stachel,
  Phys.\ Lett.\ B {\bf 675}, 312 (2009).
\bibitem{Becattini:2009fv} 
  F.~Becattini and R.~Fries,
  ``The QCD confinement transition: Hadron formation,'' in ``Relativistic heavy ion physics''
  Landolt-Bornstein {\bf 23}, 208 (2010)
  
\bibitem{Becattini:2010sk} 
  F.~Becattini, P.~Castorina, A.~Milov and H.~Satz,
  Eur.\ Phys.\ J.\ C {\bf 66}, 377 (2010).


\bibitem{Fischer:2016zzs} 
  N.~Fischer and T.~Sj\"ostrand,
  JHEP {\bf 1701}, 140 (2017).

\bibitem{Elze:1994hj} 
  H.~T.~Elze,
  Phys.\ Lett.\ B {\bf 369}, 295 (1996).
  
\bibitem{Elze:1994qa} 
  H.~T.~Elze,
  Nucl.\ Phys.\ B {\bf 436}, 213 (1995).

\bibitem{Bialas:1999zg} 
  A.~Bialas,
  Phys.\ Lett.\ B {\bf 466}, 301 (1999)

\bibitem{Akkelin:2016rhm} 
  S.~V.~Akkelin,
  Eur.\ Phys.\ J.\ A {\bf 53}, no. 12, 232 (2017).

\bibitem{Unruh:1976db} 
W.~G.~Unruh,
Phys.\ Rev.\ D {\bf 14}, 870 (1976).

\bibitem{Andersson:1983ia} 
B.~Andersson, G.~Gustafson, G.~Ingelman and T.~Sj\"ostrand,
Phys.\ Rept.\  {\bf 97}, 31 (1983).

\bibitem{Catani:1991hj} 
S.~Catani, Y.~L.~Dokshitzer, M.~Olsson, G.~Turnock and B.~R.~Webber,
Phys.\ Lett.\ B {\bf 269}, 432 (1991).

\bibitem{Frishman:2010zz} 
  Y.~Frishman and J.~Sonnenschein,
  ``Non-perturbative field theory: From two-dimensional conformal field theory to QCD in four dimensions,''
  (Cambridge University Press, Cambridge, 2010).

\bibitem{Hebenstreit:2013baa} 
  F.~Hebenstreit, J.~Berges and D.~Gelfand,
  Phys.\ Rev.\ Lett.\  {\bf 111}, 201601 (2013).

\bibitem{Casini:2009sr} 
  H.~Casini and M.~Huerta,
  J.\ Phys.\ A {\bf 42}, 504007 (2009).

\bibitem{Holzhey:1994we} 
  C.~Holzhey, F.~Larsen and F.~Wilczek,
  Nucl.\ Phys.\ B {\bf 424}, 443 (1994).
  
\bibitem{Calabrese:2004eu}
  P.~Calabrese and J.~L.~Cardy,
  J.\ Stat.\ Mech.\  {\bf 0406} (2004) P06002.

\bibitem{Calabrese:2009qy} 
  P.~Calabrese and J.~Cardy,
  J.\ Phys.\ A {\bf 42}, 504005 (2009).

\bibitem{Goykhman:2015sga} 
  M.~Goykhman,
  Phys.\ Rev.\ D {\bf 92}, no. 2, 025048 (2015).

\bibitem{BFV2}
  J.~Berges, S.~Floerchinger and R.~Venugopalan,
  arXiv:1712.09362 [hep-th].

\bibitem{tHooft:1974pnl} 
  G.~'t Hooft,
  Nucl.\ Phys.\ B {\bf 75}, 461 (1974).

\bibitem{Brandt:2016xsp} 
  B.~B.~Brandt and M.~Meineri,
  Int.\ J.\ Mod.\ Phys.\ A {\bf 31}, no. 22, 1643001 (2016).
  
\bibitem{GrosseOetringhaus:2009kz} 
  J.~F.~Grosse-Oetringhaus and K.~Reygers,
  J.\ Phys.\ G {\bf 37}, 083001 (2010).

\bibitem{Andronic:2012ut} 
  A.~Andronic, P.~Braun-Munzinger, J.~Stachel and M.~Winn,
  Phys.\ Lett.\ B {\bf 718}, 80 (2012).

\bibitem{Pal:2003rz} 
  S.~Pal and S.~Pratt,
  Phys.\ Lett.\ B {\bf 578}, 310 (2004).

\bibitem{Casini:2011kv} 
H.~Casini, M.~Huerta and R.~C.~Myers,
JHEP {\bf 1105}, 036 (2011).

\bibitem{Arias:2016nip} 
R.~Arias, D.~Blanco, H.~Casini and M.~Huerta,
Phys.\ Rev.\ D {\bf 95}, no. 6, 065005 (2017).

\bibitem{Candelas:1978gf} 
P.~Candelas and J.~S.~Dowker,
Phys.\ Rev.\ D {\bf 19}, 2902 (1979).

\bibitem{beta}
D.~N.~Zubarev, A.~V.~Prozorkevich, S.~A.~Smolyanskii, 
Theoret.\ and Math.\ Phys.\ {\bf 40}, 821 (1979);  
Ch.~G.~Van~Weert, Ann.\ Phys.\ {\bf 140}, 133 (1982);
H.~A.~Weldon, Phys.\ Rev.\ D {\bf 26}, 1394 (1982); see also
F.~Becattini, Phys.\ Rev.\ Lett.\  {\bf 108}, 244502 (2012). 

\bibitem{coldgas}
M.~Gring, M.~Kuhnert, T.~Langen, T.~Kitagawa, B.~Rauer, M.~Schreitl, I.~Mazets, D.~A.~Smith, E.~Demler, J.~Schmiedmayer
Science 337, 6100 (2012).

\bibitem{Kovner:2015hga} 
  A.~Kovner and M.~Lublinsky,
  Phys.\ Rev.\ D {\bf 92}, no. 3, 034016 (2015).

\bibitem{Kharzeev:2017qzs} 
  D.~E.~Kharzeev and E.~M.~Levin,
  Phys.\ Rev.\ D {\bf 95}, no. 11, 114008 (2017).

\bibitem{Martens:2017cvj} 
  J.~C.~Martens, J.~P.~Ralston and J.~D.~T.~Takaki,
  arXiv:1707.01638 [hep-ph].

\bibitem{Shuryak:2017phz} 
  E.~Shuryak and I.~Zahed,
  arXiv:1707.01885 [hep-ph].


\end{thebibliography}
\end{document}